\documentclass[fleqn,10pt]{wlscirep}

\usepackage{ifthen}
\usepackage{ifpdf}
\usepackage{color}

\ifpdf
\usepackage{graphicx}
\usepackage{epstopdf}
\else
\usepackage{graphicx}
\usepackage{epsfig}
\fi
\graphicspath{{./Figs/}{./}}

\usepackage{latexsym}
\usepackage{amsmath}
\usepackage{amssymb}
\usepackage{bm}
\usepackage{wasysym}

\usepackage{mathptmx}
\DeclareSymbolFont{epsilon}{OML}{ntxmi}{m}{it}
\DeclareMathSymbol{\epsilon}{\mathord}{epsilon}{"0F}

\usepackage{hyperref}

\newcommand{\amatrix}[1]{\begin{matrix} #1 \end{matrix}} 
\newcommand{\eexp}[1]{\mathrm{e}^{#1}}

\newcommand{\be}[1]{\begin{eqnarray}\ifthenelse{#1=-1}{\nonumber}{\ifthenelse{#1=0}{}{\label{e#1}}}}
\newcommand{\beq}{\begin{eqnarray}}
\newcommand{\eeq}{\end{eqnarray}} 

\newcommand{\hide}[1]{}
\newcommand{\Eq}[1]{\textcolor{blue}{{equation}\!~(\ref{#1})}} 

\newcommand{\Fig}[1]{\textcolor{blue}{Fig.}\!~\ref{#1}}
\newcommand{\sect}[1]{{\bf #1.-- }}

\renewcommand{\url}[2] {\href{#1}{[link]}}  
\newcommand{\urlprefix}{}
\newcommand{\hidea}[1]{} 

\title{Percolation, sliding, localization and relaxation in topologically closed circuits}

\author[1,*]{Daniel Hurowitz}
\author[1,*]{Doron Cohen}

\affil[1]{Department of Physics, Ben-Gurion University of the Negev, Beer-Sheva, Israel}

\hide{
The ``glassy" version of {\em random walk on disordered lattice}  
features a percolation-related crossover to variable-range-hopping,
or to sub-diffusion in one-dimension.  
The more general problem of {\em random walk in random environment}, 
where transition rates are allowed to be asymmetric, 
has been explored by Sinai, Derrida, and followers.  
It turns out that for any small amount of disorder 
an unbiased spreading in one-dimension becomes sub-diffusive,  
while for bias that exceeds a finite threshold there 
is a {\em sliding transition}, leading to a non-zero drift velocity.
We explore the implications of the percolation and sliding transitions 
for the relaxation modes of a topologically closed circuit. 
A complementary question regarding  the ``delocalization" of eigenstates 
of non-hermitian Hamiltonians has been addressed by Hatano, Nelson, and followers. 
But we show that for a conservative random-walk dynamics 
the implied spectral properties are dramatically different.
}

\begin{abstract}
Considering {\em a random walk in a random environment} in a topologically closed circuit, 
we explore the implications of the percolation and sliding transitions 
for its relaxation modes. 
A complementary question regarding  the ``delocalization" of eigenstates 
of non-hermitian Hamiltonians has been addressed by Hatano, Nelson, and followers. 
But we show that for a conservative stochastic process 
the implied spectral properties are dramatically different.
In particular we determine the threshold for under-damped relaxation, 
and observe ``complexity saturation" as the bias is increased. 
\end{abstract}

\begin{document}
\flushbottom
\maketitle
\thispagestyle{empty}

\section*{Introduction}

The original version of Einstein's Brownian motion problem
is essentially equivalent to the analysis of a simple {\em random walk}. 
The more complicated version of {\em a random walk on a disordered lattice},
features a percolation-related crossover to variable-range-hopping,
or to sub-diffusion in one-dimension~\cite{Alexander}.    
In fact it is formally like a resistor-network problem, 
and has diverse applications, e.g. 
in the context of ``glassy" electron dynamics \cite{ege,egt}. 
But more generally one has to consider Sinai's spreading problem \cite{Sinai,odh1,odh3,BouchaudReview}, 
aka {\em a random walk in a random environment}, 
where the transition rates are allowed to be asymmetric. 
It turns out that for any small amount of disorder 
an unbiased spreading in one-dimension becomes sub-diffusive,  
while for bias that exceeds a finite threshold there 
is a {\em sliding transition}, leading to a non-zero drift velocity.
The latter has relevance e.g. for studies in a biophysical context: 
population biology \cite{popbio,popbio2}, pulling pinned polymers and DNA denaturation \cite{DNA1,DNA2} 
and processive molecular motors~\cite{fisher1999force,rief2000myosin}.

The dynamics in all the above variations of the random-walk problem   
can be regarded as a stochastic process in which a particle hops from site to site.
The rate equation for the site occupation probabilities $\bm{p}  = \{p_n\}$
can be written in matrix notation as 
\be{1}
\frac{d\bm{p}}{dt} \ \ = \ \ \bm{W} \bm{p}, 
\eeq
involving a matrix ${\bm{W}}$ whose off-diagonal elements 
are the transition rates ${w_{nm}}$, 
and with diagonal elements ${-\gamma_n}$ such that each column sums to zero.
Assuming near-neighbor hopping the ${\bm{W}}$ matrix takes the form
\be{2}
\bm{W} \ \ = \ \ \left[\amatrix{
-\gamma_1   & w_{1,2}   & 0         & ... \\ 
w_{2,1}     & -\gamma_2 & w_{2,3}   & ... \\ 
0           & w_{3,2}   & -\gamma_3 & ... \\
...         & ...       & ...       & ...
}\right]
\eeq 
In Einstein's theory $\bm{W}$ is symmetric, 
and all the non-zero rates are the same.
Contrary to that, in the ``glassy" resistor-network problem (see Methods)
the rates have some distribution $P(w)$ whose small~$w$ asymptotics  is characterized 
by an exponent~$\alpha$, namely $P(w) \propto w^{\alpha-1}$ for small~$w$.
To be specific we consider  
\be{4}
P(w) \ \ = \ \ \left(\frac{\alpha}{w_c^{\alpha}}\right) w^{\alpha-1} \ (w<w_c)
\eeq
The conductivity of the network~$w_{\infty}$ is sensitive to~$\alpha$.     
It is given by the harmonic average over the $w_n$, reflecting serial addition of connectors.
It comes out non-zero in the percolating regime (${\alpha>1}$). 
For the above distribution ${w_{\infty}=[(\alpha{-}1)/\alpha]w_c}$.    

In Sinai's spreading problem $\bm{W}$ is allowed to be asymmetric.
Accordingly the rates at the $n$th bond can be written 
as $w_n\eexp{\pm\mathcal{E}_n/2}$ 
for forward  and backward transitions respectively. 
For the purpose of presentation we assume that the stochastic field~$\mathcal{E}$  
is box distributed within ${[s-\sigma,s+\sigma]}$.
We refer to~$s$ as the bias:  
it is the pulling force in the case of depinning polymers and DNA denaturation; 
or the convective flow of bacteria relative to the nutrients in the case of population biology; 
or the affinity of the chemical cycle in the case of molecular motors.

Our interest is in the relaxation dynamics 
of finite $N$-site ring-shaped circuits \cite{nef,nes}, 
that are described by the stochastic equation \Eq{e1}.
The ring is characterized by its so-called affinity, 
\be{18}
S_{\circlearrowleft} \ \ \equiv \ \  \sum_{n=1}^{N} \mathcal{E}_n  \ \ \equiv \ \  N \, s
\eeq 
The $N$~sites might be physical locations in some lattice 
structure, or can represent steps of some chemical-cycle. 
For example, in the Brownian motor context $N$~is the number 
of chemical-reactions required to advance the motor one pace. 
We are inspired by the study of of non-Hermitian quantum Hamiltonians 
with regard to vortex depinning in type II superconductors \cite{Hatano1,Hatano2,Shnerb1};   
molecular motors with {\em finite} processivity \cite{brm1,brm2}; 
and related works \cite{Brouwer,Goldsheid,Zee}.
In the first example the bias is the applied transverse magnetic field;  
and $N$~is the number of defects to which the magnetic vortex can pin.
In both examples conservation of probability is violated.

\section*{Scope} 

\begin{figure}

\centering

\includegraphics[height=3.7cm]{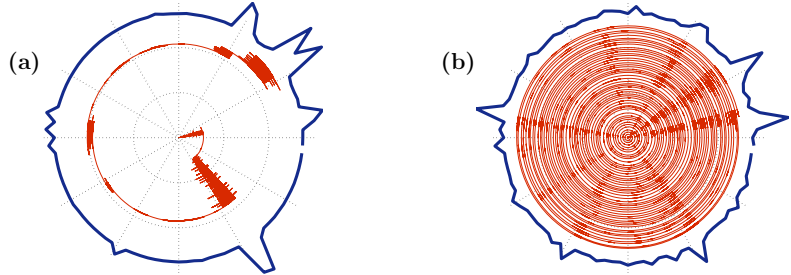}

\caption{\label{traj}
{\bf Simulated trajectory of a particle on a disordered-ring.}
The number of sites is $N{=}100$ and the disorder strength is $\sigma{=}5$.
The radial direction is time and the angle is the position. 
(a) For small affinity ($s{=}0.88$) the dynamics is over-damped. 
(b) For large affinity ($s{=}2.97$) the dynamics is under-damped. 
The outer thick line is the steady-state distribution (see Methods). 
}
\end{figure}

\begin{figure}

\includegraphics[height=9cm]{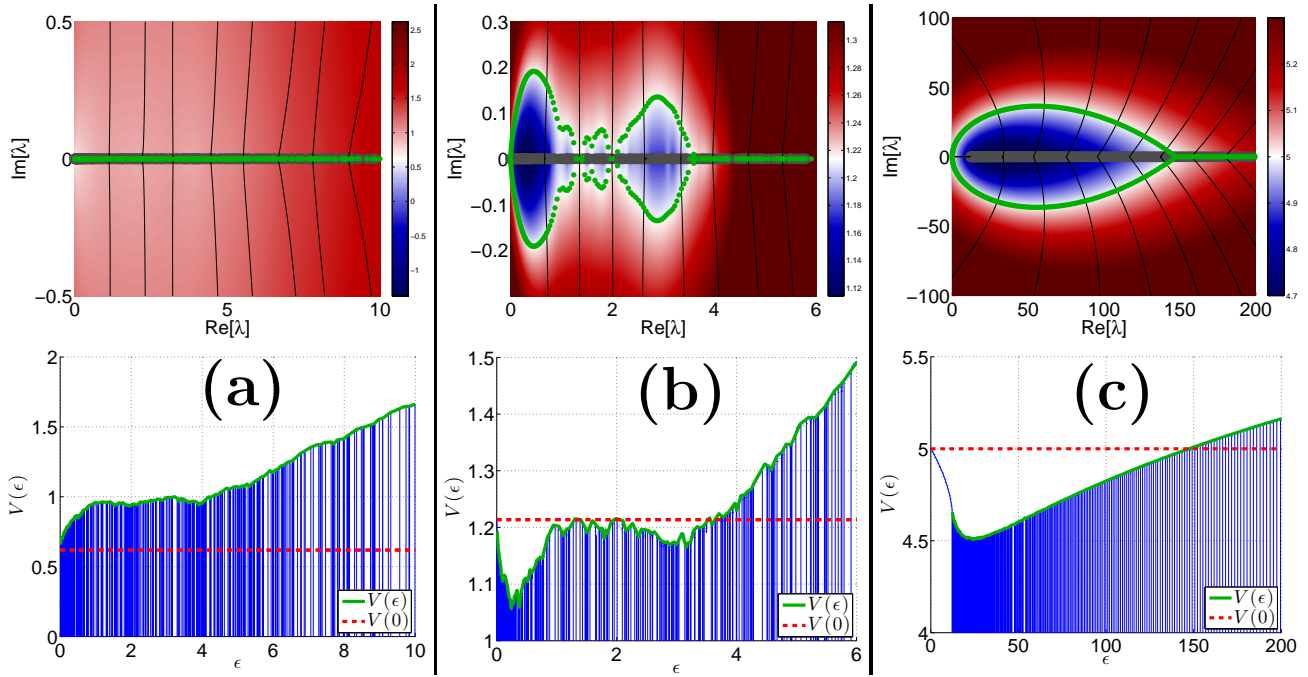}

\caption{\label{froots} 
{\bf The emergence of complexity in the relaxation spectrum.} 
The upper plot in each panel displays a representative example for a relaxation spectrum.
The $\{\lambda_k\}$ are indicated by green points in the complex plane. 
The ring has $N{=}500$ sites. The disorder is $\sigma{=}5$, 
hence the calculated threshold values are ${s_{1/2}=1.77}$ and ${s_1=2.7}$ and ${s_{\infty}=5}$.
In (a) the affinity is ${s = 1.24 < s_{1/2}}$ and the spectrum is real.
In (b) the affinity is ${s = 2.43}$ and the spectrum has   
several complex bubbles separated by real segments. 
In (c) the affinity is ${s = 10 > s_{\infty}}$ and the real spectrum has a gap, 
while the complex spectrum is a fully developed complex bubble, 
tangent to the origin (no gap). 
The ``electrostatic field" that is associated with the characteristic equation 
is represented by a few field-lines, while the background color provides 
visualization of the corresponding electrostatic potential. 
The spectrum is obtained by looking  for the intersections 
of the field lines with the equipotential line ${V(z)=V(0)}$ 
that goes through the origin (indicated in white). 
The lower panels plot the potential $V(\epsilon)$ along the real axis. 
The horizontal dashed line is~$V(0)$. 
}
\end{figure}

In this article we report how the spectral properties of the matrix $\bm{W}$ 
depend on the parameters $(\alpha,\sigma,s)$, as defined above. 
These parameters describe respectively the resistor-network disorder, 
the stochastic-field disorder, and the average bias field.    
The eigenvalues $\{-\lambda_k\}$ of $\bm{W}$ are associated 
with the relaxation modes of the system. 
Due to conservation of probability ${\lambda_0=0}$, 
while all the other eigenvalues ${\{\lambda_k\}}$ have positive 
real part, and may have an imaginary part as well. 
Complex eigenvalues imply that the relaxation is not over-damped: 
one would be able to observe an oscillating density during relaxation, 
as demonstrated in \Fig{traj}.    
The panels of \Fig{froots} provide some representative spectra.
As the bias~$s$ is increased a complex bubble appears at the bottom 
of the band, implying delocalization of the eigenstates.   
Our results for the complexity threshold~$S_c$ are summarized 
in \textcolor{blue}{Table} \ref{tbl}, and demonstrated in \Fig{figSc}. 
The number of complex eigenvalues grows as a function 
of the bias, as demonstrated in \Fig{figCplxSat},  
but asymptotically only a finite fraction of the spectrum becomes complex. 
Our objective below is to explain analytically the 
peculiarities of this delocalization transition, 
to explain how it is affected by the percolation and by the sliding thresholds, 
and to analyze the complexity-saturation effect.  

{\em Note about semantics:} What we called above a ``percolation-like transition" at ${\alpha=1}$ 
means that for an infinite chain, in the statistical sense, the conductivity ($w_{\infty}$) 
is zero for ${\alpha<1}$ and becomes non-zero for ${\alpha>1}$. 
Clearly, if the bond distribution~$P(w)$ were bi-modal (if the $w_n$ were zeros or ones), 
we would not have in one-dimension a percolation transition~\cite{Saberi20151}.

\begin{table}

\centering

\begin{tabular}{|c|cc|c|c|}
\hline 
Type of disorder & Parameters & &  $S_c$ for large $N$ & Remarks \\ 
\hline
Resistor-network & $\alpha{<}\frac{1}{2}$  & $\sigma{=}0$    &  $S_c = \infty$ &  non-percolating (``disconnected ring") \\
Resistor-network & $\frac{1}{2}{<}\alpha{\ll}1$ & $\sigma{=}0$   &  $S_c \sim \mathcal{O}(1)$ & residual percolation (``weak link") \\
Resistor-network & $\alpha{>}1$ & $\sigma{=}0$   &  $S_c \propto 1/\sqrt{N}   $ & percolating (conductivity $w_{\infty}{>}0$) \\
Stochastic field & $\alpha{>}\frac{1}{2}$  & $\sigma{>}0$ & $S_c \approx N \, s_{1/2}$ & lower than sliding threshold at $Ns_1$ \\
\hline
\end{tabular}

\caption{\label{tbl}
{\bf The complexity threshold for different types of disorder} (aka delocalization transition). 
We distinguish between resistor network disorder (${\alpha<\infty}$) and stochastic field disorder (${\sigma>0}$).  
The threshold $s_{1/2}$ is determined by the the latter. 
It is smaller than the $s_1$~threshold of the sliding transition.    
Note that the thresholds $s_{\mu}$ depend neither on~$N$ nor on~$\alpha$. 
} 

\end{table}

\section*{Stochastic spreading}

We first consider an opened ring, namely a disordered chain. 
The asymmetry can be gauged away, and $\bm{W}$ becomes similar 
to a symmetric matrix~$\bm{H}$ (see Methods). 
The statistics of  the off-diagonal elements of~${\bm H}$
is characterized by~$\alpha$, 
while the statistics of the diagonal elements 
is affected by~$\sigma$ and~$s$ too.   
The eigenvalues $\{-\epsilon_k\}$ of $\bm{H}$ are real.
In the absence of disorder they form a band ${[\epsilon_s,\epsilon_{\infty}]}$
where ${\epsilon_{s,\infty}=2[\cosh(s/2)\mp1]}$. 
%
%
If the stochastic-field disorder has a Gaussian statistics 
the gap ${[0,\epsilon_s]}$ is closed \cite{odh3}. 
In this case there is an analytical expression 
for the spectral density in terms of Bessel functions.
The expression features 
\be{3}
\rho(\epsilon) \ \propto \ \epsilon^{\mu-1} \ \ \ \ \text{(for small $\epsilon$)}
\eeq
with no gap. The exponent is related to the bias via ${s=(1/2) \sigma^2 \mu}$. 
In the present work we assume the more physically appealing log-box 
disorder for which the relation between $s$ and $\mu$ is as follows (see Methods):
\be{5}
s_{\mu} \ \ = \ \ \frac{1}{\mu} \ln\left( \frac{\sinh (\sigma\mu)}{\sigma\mu} \right)
\eeq
Unlike Gaussian disorder the range of possible rates is bounded,  
and we see that ${s_{\infty}=\sigma}$ is finite rather than infinite.
For ${s>s_{\infty}}$ a gap opens up, meaning that $\epsilon_s$ acquires a finite non-zero value.

In order to have a non-zero drift velocity along an {\em infinite} chain 
two conditions have to be satisfied: First of all the system has 
to be percolating (${\alpha>1}$) such that its conductivity $w_{\infty}$ is non-zero; 
Additionally one requires the bias~$s$ to exceed the threshold $s_1$, 
such that ${\mu>1}$. This is known as the ``sliding transition". One obtains  
\beq
v_{\text{drift}} \ \ = \ \ \eexp{\frac{1}{2}(s_1-s_{1/2})} \left[ 2 \sinh\left(\frac{s-s_1}{2}\right) \right] w_{\infty}
\hspace{2cm} \text{[sliding regime ($s>s_1$)]}
\eeq
Contrary to that, in the regime ${s<s_1}$ there is a build-up of an activation-barrier 
that diverges in the ${N\to\infty}$ limit, hence the drift velocity vanishes.  
The above mentioned spectral properties imply that for ${\mu<1}$  
the spreading of a distribution along an infinite chain 
becomes anomalously slow and goes like $x\sim t^{\mu}$.  
Concerning the second moment: for ${\mu<1/2}$ 
the diffusion coefficient is zero, reflecting sub-diffusive spreading. 
In the absence of bias ($\mu \to 0$) the spreading becomes logarithmically slow.

The absence of resistor-network-disorder formally 
corresponds to $\alpha=\infty$ in \Eq{e4}, 
meaning that all the $w_n$ have the same value.
The introduction of resistor-network-disorder ($\alpha<\infty$) 
modifies the spectral density \Eq{e3} at higher energies 
(see \Fig{fdos} of the Methods for illustration).
In the absence of bias, for ${\alpha>1}$, 
the continuum-limit approximation features ${\mu_{\alpha}=1/2}$.
This reflects a normal diffusive behavior 
as in Einstein's theory of Brownian motion. 
Below the percolation threshold, namely for ${\alpha<1}$,
normal diffusion is suppressed~\cite{Alexander}, 
and the spectral exponent becomes ${\mu_{\alpha}=\alpha/(1{+}\alpha)<1/2}$. 
In the other extreme of very large bias, 
the diagonal disorder in~$\bm{H}$ dominates,  
leading to trivially localized eigenstates. 
Hence for very large bias we simply have ${\mu=\alpha}$
irrespective of the percolation aspect.  

The conclusion of this section requires a conjecture that is supported 
by our numerical experience (we are not aware of a rigorous derivation):  
As the bias $s$ is increased, the exponent $\mu$ becomes larger,
as implied by \Eq{e5}, but it cannot become larger than~$\alpha$.
We shall use this conjecture in order 
to explain the observed implications of 
resistor-network-disorder.

\section*{Relaxation}

We close an $N$-site chain into a ring and wonder 
what are the relaxation modes of the system. 
The starting point of our analysis is the characteristic
equation for the eigenvalues of $\bm{W}$.
Assuming that we already know what are the eigenvalues 
of the associated symmetric matrix $\bm{H}$,    
the characteristic equation takes the form \cite{det1} (see Methods)
\be{20}
\prod_k \left(\frac{z-\epsilon_k(s)}{\overline{w}}\right) 
\ \ = \ \ 2\left[\cosh\left(\frac{S_{\circlearrowleft}}{2}\right)-1\right] \, (-1)^N 
\eeq
where ${\overline{w}}$ is the geometric average of all the rates.
The bias~$s$ affects both the $\epsilon_k$ and the right hand side.
This equation has been analyzed in \cite{Shnerb1} 
in the case of a non-conservative matrix~$\bm{W}$ 
whose diagonal elements $\gamma_n$ are {\em fixed}, 
hence the $\epsilon_k(s)$ there do not depend on~$s$. 
Consequently, as~$s$ of \Eq{e18} is increased beyond 
a threshold value $s_{c}$, the eigenvalues in the middle 
of the spectrum become complex. 
As~$s$ is further increased beyond some higher threshold value, 
the entire spectrum becomes complex. 
As already stated in the introduction, this is not the scenario 
that is observed for our conservative model.
Furthermore we want to clarify how the percolation 
and sliding thresholds are reflected.

Already at this stage one should be aware of the immediate 
implications of conservativity. First of all ${z=\lambda_0=0}$ 
should be a root of the characteristic equation. 
The associated eigenstate is the non-equilibrium steady state (NESS), 
which is an extended state (see Methods). 
In fact it follows that the localization length has 
to diverge as ${\lambda\rightarrow0}$. 
This is in essence the difference between the 
conventional Anderson model (Lifshitz tails at the band floor) 
and the Debye model (phonons at the band floor). 
It is the latter picture that applies in the 
case of a conservative model.

\section*{Electrostatic picture}

In order to gain insight into the characteristic equation we 
define an ``electrostatic" potential by taking the log 
of the left hand side of \Eq{e20}. Namely, 
\be{22}
\Psi(z) \ \ = \ \ \sum_k \ln\left(z-\epsilon_k\right) \ \ \equiv \ \ V(x,y)+iA(x,y)
\eeq
where ${z=x+iy}$, and for simplicity of presentation we set here and below 
the units of time such that ${\overline{w}=1}$.  
The constant ${V(x,y)}$ curves correspond to potential contours,
and the constant ${A(x,y)}$ curves corresponds 
to stream lines. The derivative $\Psi'(z)$ corresponds to the field, 
which can be regarded as either electric or magnetic field up to a 90deg rotation.       
Using this language, the characteristic equation \Eq{e20} takes the form
\be{21}
V(x,y)=V(0); \ \ \ \ A(x,y)=2\pi*\text{integer} 
\eeq
Namely the roots are the intersection of the field lines with the 
potential contour that goes through the origin (\Fig{froots}). 
We want to find what are the conditions for getting 
a real spectrum from \Eq{e21}, and in particular what 
is the threshold $s_c$ for getting complex eigenvalues 
at the bottom of the spectrum. 
We first look on the potential along the real axis:
\beq
V(\epsilon) \ \ = \ \  \int \ln \left(|\epsilon-x' \right|) \rho(x')dx' 
\eeq
In regions where the $\{\epsilon_k\}$ form a quasi-continuum,  
one can identify $(1/N)V(\epsilon)$ as the Thouless expression  
for the inverse localization length \cite{Shnerb1}.
The explicit value of $V(0)$ is implied by \Eq{e20}, 
namely ${V(0)=\ln[2(\cosh(S_{\circlearrowleft}/2)-1)]}$.   
For a charge-density that is given by \Eq{e3}, with some cutoff $\epsilon_c$,
the derivative of the electrostatic potential at the origin is (see Methods)
\be{19}
V'(\epsilon) \ \ \approx \ \ \frac{ \epsilon^{\mu-1}}{\epsilon_c^{\mu}} \pi \mu \cot(\pi \mu)
\eeq
One observes that the sign of $V'(\epsilon)$ is positive for ${\mu<1/2}$, 
and negative for ${\mu>1/2}$.
Some examples are illustrated in \Fig{froots}.
Clearly, if the envelope of $V(\epsilon)$ is above 
the ${V=V(0)}$ line, then the spectrum is real, and the $\lambda_k$ are roughly 
the same as the $\epsilon_k$, shifted a bit to the left.

From the above it follows that the threshold~$s_c$ 
for the appearance of a complex quasi-continuum 
is either  ${V(\epsilon_s)<V(0)}$  or  ${V'(0)<0}$, 
depending on whether $\varrho(\epsilon)$ is gapped or not. 
In the latter case it follows from \Eq{e19} that ${s_c=s_{1/2}}$.
We note that for the Gaussian model of \cite{odh3}
one obtains ${V(\epsilon \rightarrow \infty) = \text{const}}$,  
implying that the entire spectrum would go from real to complex at $s=s_{1/2}$. 
In general this is not the case: the complex spectrum typically 
forms a ``bubble" tangent to the origin, or possibly 
one may find some additional bubbles as in \Fig{froots}b (upper plot).

\section*{Resistor-network disorder}

The prediction ${S_c = Ns_{1/2}}$ 
assumes full stochastic-field disorder over the whole ring.
One may have the impression that this result suggests ${S_c=0}$ 
in the absence of stochastic field disorder, 
because ${s_{1/2}=0}$ for $\sigma{=}0$.    
We shall argue below that this is a false statement.
Clearly the prediction ${S_c = Ns_{1/2}}$  is irrelevant
if one link is disconnected (say ${w_1=0}$). 
In the latter case one would expect ${S_c=\infty}$. 
Naively an infinite $S_c$ might be expected 
throughout the non-percolating regime (${\alpha<1}$). 
But we shall argue that this is a false statement too.

Consider first a clean ring. Recall that it features a continuous 
spectral density $\varrho(\epsilon)$ that is supported 
by ${[\epsilon_s,\epsilon_{\infty}]}$. 
An isolated defected bond  contributes an isolated eigenvalue 
outside of the band. This is like having an impurity.
A detailed example for this, is presented in the Methods section,  
where we establish that for a weak-link ${S_c}$ is finite, 
and independent of~$N$. Similar analysis can be carried-out for other 
types of isolated defects. 
 
\begin{figure}

\centering

\includegraphics[height=6cm]{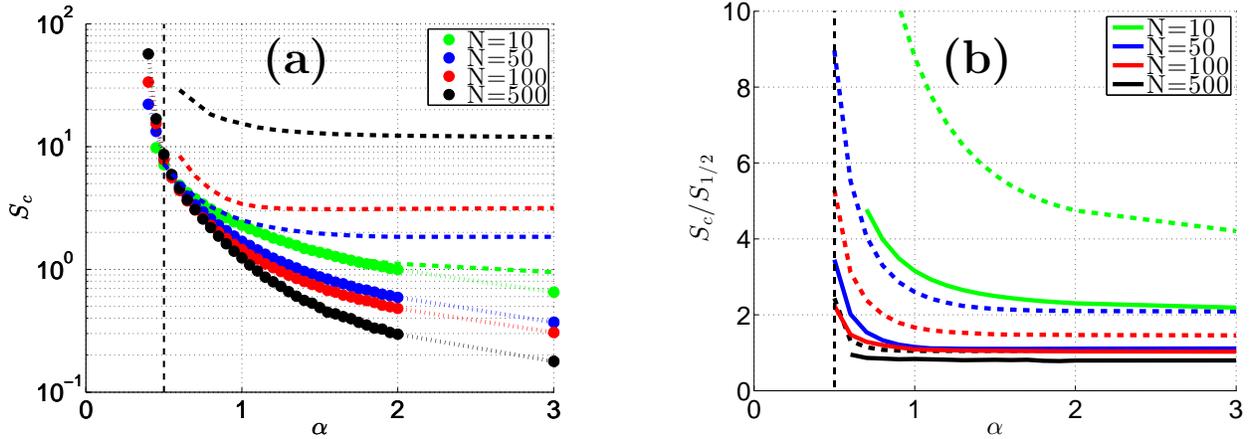}

\caption{\label{figSc}
{\bf The complexity threshold.}
We plot $S_c$ versus~$\alpha$ for rings with ${N=10,50,100,500}$ sites, 
and representative values of stochastic field disorder.
The dotted, dashed and solid lines are for $\sigma=0,0.5,1.0$ respectively.
Average has been taken over $\sim100$ realizations for 
each data point. 
In panel~(a) the data points of the $\sigma=0$ curve are indicated by circles. 
We clearly see that the effective percolation threshold, 
beyond which $S_c$ becomes finite, is ${\alpha=1/2}$ rather than ${\alpha=1}$.
For a percolating disorder $S_c$ diminishes as~$N$ is increased,      
while for marginal percolation (${1/2<\alpha\ll 1}$)  
the threshold becomes $N$~independent as for a clean-ring 
that has a single weak-link.
The dashed lines in panel~(a) are for ${\sigma=0.5}$.
In order to demonstrate that they agree 
with $S_c = Ns_{1/2}$ we plot additional curves 
for ${\sigma=1.0}$ in panel~(b), 
and scale the vertical axis appropriately. 
Comparing the two panels we see that the $s_{1/2}$ based prediction becomes 
valid once it exceeds the resistor-network threshold.   
}
\end{figure}

Full resistor-network disorder (${\alpha<\infty}$ with ${\sigma=0}$) can be regarded 
as having some distribution of ``weak-links" along the ring. 
We can speculate that for large~$N$  
there are two limits: either $S_c \to \infty$ 
or ${S_c \to 0}$ depending on whether 
the ring is percolating or not. 
Our numerical results are presented in \Fig{figSc}a.
Surprisingly the effective percolation threshold 
is not ${\alpha=1}$, but ${\alpha=1/2}$.
The threshold $S_c$ becomes infinite only if ${\alpha<1/2}$. 
We are able to predict this numerical observation  
using the electrostatic picture:
In the regime ${\alpha<1/2}$ the spectral density $\varrho(\epsilon)$ 
is characterized by a an exponent ${\mu<1/2}$. 
Namely it goes from ${\mu=\alpha/(1{+}\alpha)<1/2}$ 
for small $\epsilon$, to~${\mu=\alpha<1/2}$ for large~$\epsilon$.  
Consequently $V(\epsilon)$ becomes a monotonic increasing 
function, and it follows from the the reasoning of the 
previous section that all eigenvalues are real.

For a percolating but disordered resistor-network (${1<\alpha<\infty}$ but ${\sigma=0}$)
we expect ${S_c \propto 1/N^{1/2}}$, see Methods.    
The marginally-percolating regime ($1/2<\alpha<1$) is conceptually 
like having sparsely distributed weak-links. 
Accordingly, as ${\alpha \to 1/2}$ the threshold ${S_c}$ becomes independent of~$N$.
These predictions are confirmed numerically in \Fig{figSc}a.
The additional numerical results that are presented in \Fig{figSc}b
demonstrate what happens if we add stochastic field disorder: 
The prediction ${S_c = Ns_{1/2}}$ becomes valid 
once it exceeds the resistor-network threshold.

\begin{figure}

\centering

\includegraphics[height=5.4cm]{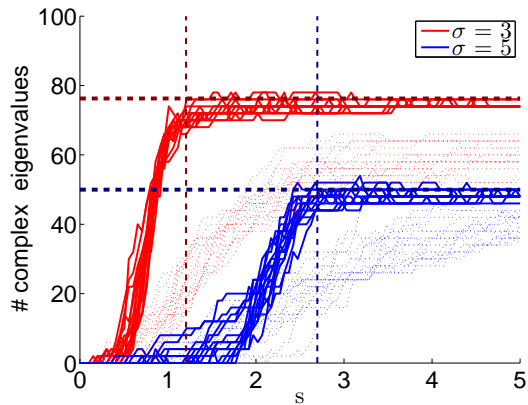}

\caption{\label{figCplxSat}
{\bf Complexity saturation.}
The number of complex eigenvalues is counted for a ring with $N{=}100$ sites, 
for various values of the affinity~$s$. Each red line corresponds to a different
realization of field disorder with $\sigma{=}3$ (red) and $\sigma{=}5$ (blue). 
The vertical lines are the corresponding values of~$s_1$, 
at which the sliding transition occurs. 
We see that the asymptotic fraction of complex eigenvalues saturates. 
The horizontal dashed line are the analytical estimates of \Eq{e101}. 
If the lattice were continuous with Gaussian disorder, 
the number of complex eigenvalues would go to~$100\%$.
In the background a disordered resistor network with ${\alpha=0.9}$ is shown. 
The crossover is blurred and the saturation value is lower compared to \Eq{e101}.  
}
\end{figure}

\section*{Complexity saturation}

The characteristic equation for the eigenvalues is given by \Eq{e20}.
In the nonconservative case, the eigenvalues of $\bm{H}$ do not depend on $s$,
thus raising $s$ will eventually make the entire spectrum complex. 
For a conservative matrix, however, $V(\epsilon)$ is also a function of $s$, 
so increasing $s$ raises $V(\epsilon)$  at the same rate.
Taking $s$ to be as large as desired, the eigenvalues of~$H$ 
become trivially ${\epsilon_n \approx \gamma_{n} \approx w_n \eexp{\mathcal{E}_n/2}}$, 
and the equation $V(\epsilon)=V(0)$ for the upper cutoff $\epsilon_c$ 
of the complex energies takes the form
\be{24}
\overline{\ln\left[ \epsilon - w\eexp{\mathcal{E}/2} \right]} \ \ = \ \ s/2 
\eeq
It is natural to write the stochastic field as $\mathcal{E}=s+\varsigma$, 
such that ${\varsigma\in[-\sigma,+\sigma]}$. 
For the purpose of presentation we assume that~$w{=}1$.  
Then the spectrum stretches from $\epsilon_s=\eexp{(s-\sigma)/2}$ 
to ${\epsilon_c=\eexp{(s+\sigma_c)/2}}$,  
where $\sigma_c$ is the solution of
\beq
\int_{-\sigma}^{\sigma} \ln \left| \eexp{{\varsigma}/2} - \eexp{\sigma_c /2}\right| d\varsigma  \ \ = \ \ 0 
\eeq
It follows that the fraction of complex eigenvalues is 
\be{101}
\text{fraction} 
= \frac{1}{N}\int_{\epsilon_s}^{\epsilon_c} \rho(\epsilon) d\epsilon
= \frac{1}{\sigma}\ln\left( \frac{\epsilon_c}{\epsilon_s} \right) 
= \frac{\sigma_c + \sigma}{2\sigma}
\eeq
We demonstrate the agreement with this formula in \Fig{figCplxSat}.
We plot there also what happens if resistor-network disorder is introduced.
We see that for small~$\alpha$ the crossover is not as sharp and 
the saturation value is lower than \Eq{e101} as expected from \Eq{e24}.

\section*{Discussion} 

We have shown that the relaxation properties of a closed circuit (or chemical-cycle), 
whose dynamics is generated by a conservative rate-equation,  
is dramatically different from that of a biased non-hermitian Hamiltonian. 
The transition to complexity (under-damped dynamics, see \Fig{traj}b) 
depends on the type of disorder as summarized in \textcolor{blue}{Table} \ref{tbl}. 
Surprisingly it happens at ${\alpha=1/2}$ before the ${\alpha{=}1}$ percolation transition, 
and at ${\mu=1/2}$ before the ${\mu{=}1}$ sliding transition.
Further increasing the bias does not lead to full delocalization, instead a ``complexity saturation" is observed.

In our analysis, we were able to bridge between the works of Hatano, Nelson, Shnerb, and followers, 
regarding the spectrum of non-hermitian Hamiltonians; the works of Sinai, Derrida, and followers, 
regarding random walks in random environments; and the works of Alexander and co-workers 
regarding the percolation related transition in ``glassy" resistor network systems.
Furthermore we have uncovered a related misconception concerning processive  molecular 
motors, contradicting a widespread conjecture regarding the equivalence 
to a uni-directional hopping model with a broad distribution of dwell times (see below). 

Spreading processes in disordered systems have been widely studied.
In the pioneering work of Derrida \cite{odh1}, the velocity and diffusion coefficient 
in the steady state were found by solving the rate equation for an $N$-site periodic lattice,  
and then taking the limit ${N \to \infty}$.
In \cite{odh2,odh3} the same results have been obtained using a Green function method, 
which requires averaging over realizations of disorder 
rather than considering a periodic chain. 
The main shortcoming of both approaches is that going beyond the steady-state is very difficult. 
Yet another approach, followed by Kafri, Lubensky and Nelson \cite{brm2}, 
is to utilize the perspective of Hatano, Nelson and Shnerb \cite{Hatano1,Hatano2,Shnerb1}, 
who studied the entire spectrum by diagonalization of the pertinent non-hermitian matrix.
This method is especially appealing if one is interested in the behavior of the system at long times. 
Additionally, this method accounts for a closed topology, an aspect disregarded in the other methods. 

In this context we would like to highlight the study of the long-time behavior of processive  molecular motors, 
such as RNAp or DNAp, moving along heterogeneous DNA tracks \cite{brm2}. 
The so-called {\em stall force} of the motor corresponds to a bias given by~$s_1$. 
For ${s<s_1}$ the drift becomes anomalous.
The common wisdom was that the relaxation spectrum remains complex for any~$s$,  
but with an anomalous density for ${s<s_1}$. 
This statement had been supported by a conjectured equivalence to a uni-directional hopping model   
with a broad distribution of dwell times, that has been proposed 
by Bouchaud, Comtet, Georges, and Le Doussal \cite{odh3}. 
Thus it has been concluded that {\em reality} requires finite-processivity.
In the present work we have established that the reality of the spectrum (over-damped relaxation)
prevails also for infinite-processivity, but the threshold is~$s_{1/2}$ 
rather than~$s_1$. Furthermore we have provided insights regrading 
various ingredients that affect the breakdown of reality; the emergence 
of complexity; and its ultimate saturation.


\clearpage

\section*{Methods}  


\sect{The percolation threshold}
An example where the percolation issue arises is provided by 
the analysis of relaxation in ``glassy" networks \cite{ege,egt}, 
where the sites are distributed randomly in space, 
and the rates depend exponentially on the inter-site distance, 
namely ${w \propto \exp(-r/\xi)}$. 
In such type of model there is a percolation-related crossover 
to variable-range-hopping \cite{pts}. But in one-dimension
there is a more dramatic crossover to sub-diffusion \cite{Alexander}.
The statistics of the inter-site distances is Poisson ${\text{Prob}(r) \propto \exp(-r/a)}$, 
where $a$ is the mean spacing, and therefore ${\alpha=\xi/a}$ in \Eq{e5}.     
The diffusion coefficient is the harmonic average over~$w_n$, 
reflecting serial addition of connectors. It becomes zero for ${\alpha<1}$. 

\begin{figure}[b!]

\centering

\includegraphics[height=6cm]{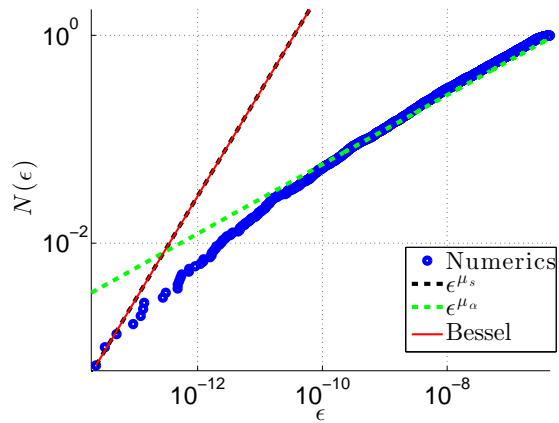}

\caption{\label{fdos}
{\bf The spectrum of the associated hermitian matrix.}
We calculate numerically the integrated density, 
which counts the eigenvalues ${\{\epsilon_k<\epsilon\}}$ of $\bm{H}$ 
for a ring with $N{=}3000$ sites. 
The system is characterized by a percolation exponent ${\mu=\mu_{\alpha}=1/3}$, 
and by a scaled affinity ${\mu=\mu_s=1}$. 
The stochastic-field distribution is with $\sigma{=}2$. 
The blue points are are results of numerical diagonalization. 
There is a crossover from density that corresponds to $\mu_s$ (dashed black line), 
to density that corresponds to $\mu_{\alpha}$ (dashed green line). 
The red line is the Bessel expression of \cite{odh3}. 
}
\end{figure}

The percolation control-parameter~${\alpha}$ 
is reflected in the exponent~$\mu$ that characterizes 
the spectral function \Eq{e3}.
As explained in the main text,  
the exponent~$\mu$ is further affected by the bias~$s$.  
See \Fig{fdos} for illustration. 
\\

\sect{The NESS formula}
Following the derivation in \cite{nef} the explicit formula for the NESS is   
\beq
p_n \ \propto \ \left( \frac{1}{w_{\overrightarrow{n}}}\right)_s e^{-\left(U(n)-U_s(n)\right)}
\eeq
where $U(n)$ is the stochastic potential that is associated with the stochastic field 
such that ${\mathcal{E}_n = U(n)-U(n{-}1)}$.
The transitions in the drift-wise direction are $w_{\overrightarrow{n}}=w_n\eexp{\mathcal{E}_n/2}$, 
and the subscript~$s$ indicates drift-wise smoothing over a length scale~$1/s$. 
In the absence of bias the smoothed functions are constant and we get the canonical equilibrium state.    
\\

\sect{Handling $\bm{W}$}
Define the diagonal matrix $\bm{U}=\mbox{diag}\{U(n)\}$. 
The stochastic field can be made uniform, as in \cite{Shnerb1}, 
by performing a similarity transformation ${\tilde{\bm{W}} = \eexp{{\bm{U}/2}} \bm{W} \eexp{-{\bm{U}/2}}}$,  
leading to 
\be{16}
\tilde{\bm{W}} \ = \ 
\text{diagonal}\Big\{-\gamma_{n}\Big\} 
+\text{offdiagonal}\Big\{  w_{n}\eexp{\pm \frac{S_{\circlearrowleft}}{2N}}  \Big\}
\eeq
where the "$\pm$" are for the forward and backward transitions respectively.
Note that the $s$-dependent statistics of the~$\mathcal{E}_n$ is still hiding in the diagonal elements.
The associated symmetric matrix $\bm{H}$ is defined by setting ${S_{\circlearrowleft}=0}$.
Then one can define an associated spectrum ${\{-\epsilon_k\}}$.
For an open chain setting ${S_{\circlearrowleft}=0}$ can be regarded 
as a gauge transformation of an imaginary vector potential.
For a closed ring ${S_{\circlearrowleft}}$ is like an 
imaginary Aharonov-Bohm flux, and cannot be gauged away. 
For the spectral determinant the following expression is available \cite{det1}:     
\be{161}
\det (z+\bm{W})  
\ \ = \ \  \det (z+\tilde{\bm{W}}) 
\ \ = \ \ \det (z+{\bm{H}}) \ - \ 2\left[\cosh\left(\frac{{S_{\circlearrowleft}}}{2}\right)-1\right] \prod_{n=1}^N (-w_n)
\eeq
This leads to the characteristic equation, \Eq{e20}.
\\

\sect{Finding $s_{\mu}$}
The cummulant generating function of the stochastic field 
can be written as $g(\mu)=(s-s_{\mu})\mu$,  
where the $s_{\mu}$ are defined via the following expression:   
\be{362}
\left\langle  e^{-\mu\mathcal{E}}\right\rangle \ \ \equiv \ \ \eexp{-(s-s_{\mu})\mu} 
\eeq
If the stochastic field has normal distribution 
with standard deviation $\sigma$, then ${s_{\mu}=(1/2) \sigma^2 \mu}$.
For our log-box distribution \Eq{e5} applies. 
The finite value of $s_{\infty}$ reflects that $\mathcal{E}$ is bounded.  
\\

\sect{Finding $V'(0)$}
To derive \Eq{e19} we assume an integrated density of states that corresponds to \Eq{e3}, 
namely, ${\mathcal{N}(\epsilon) = (\epsilon/\epsilon_c)^{\mu}}$, 
where $\epsilon_c$ is some cutoff that reflects the discreteness of the lattice.
After integration by parts the electrostatic potential along the real axis is given by 
\beq
V(\epsilon) \ \ = \ \ -  \int_0^{\epsilon_c} \frac{\mathcal{N}(x)}{x-\epsilon}dx
\eeq
While calculating the derivative we assume ${\epsilon \ll \epsilon_c}$, 
hence taking the upper limit of the scaled integral as infinity: 
\beq
V'(\epsilon) \ \ &=& \ \ 
\frac{\mu}{\epsilon_c^{\mu}} \epsilon^{\mu-1} \int_0^{\infty} \frac{z^{\mu-1}}{z-1}dz 
\\
&=&   
-\frac{\mu}{\epsilon_c^{\mu}}  \epsilon^{\mu-1} \ B_{\infty}(\mu,0)
\eeq  
where $B_u(a,b)$ is the Incomplete Euler Beta function.
Taking the Cauchy principal part we get
\beq
B_{\infty}(\mu,0) &=&  
\lim_{\delta \to 0}  \left[ B_{1-\delta}(\mu,0) - B_{1-\delta}(1-\mu,0) \right]  
\\
&=& \psi(1-\mu)-\psi(\mu) = \pi \cot(\pi \mu)
\eeq
where $\psi(z)$ is the digamma function,
and the last equality has been obtained by the reflection formula. 
\\

\begin{figure}

\centering

\includegraphics[height=6.5cm]{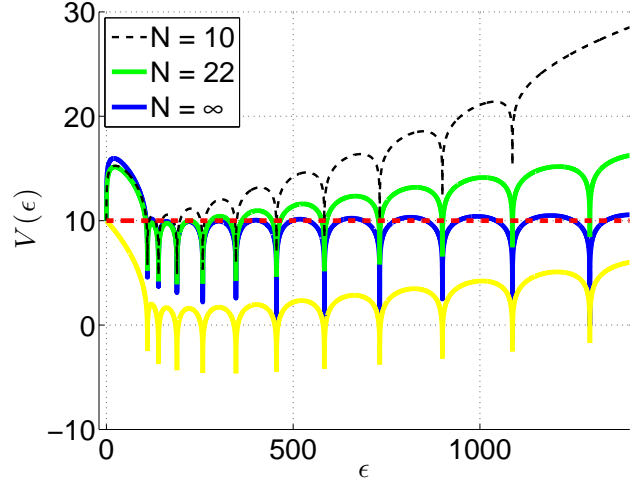}

\caption{\label{figReconstruction} 
{\bf Graphical illustration of the the characteristic equation for a ring with a weak link.} 
The red line is $V(0)$.  
The blue line is $V(\epsilon)$ as deduced from the LHS of \Eq{e50}
with ${L=1}$, and ${g=10^{-3}}$ and ${S_{\circlearrowleft}=20}$.
The yellow line is an attempted reconstruction of $V(\epsilon)$ 
from the first $N=22$ roots $\epsilon_k$ of the $S_{\circlearrowleft}{=}0$ equation. 
The green line is a proper reconstruction that takes into account 
an impurity term $\epsilon_0$. The deviation from the blue line 
for large~$k$ is due to finite truncation: compare the ${N{=}10}$ line with the ${N{=}22}$ line. 
}
\end{figure}

\hide{
\sect{Finding $S_c$ due to a biased-link}
We consider a clean ring. We assume that the stochastic field 
over one bond is exceptionally large compared to all other bonds.   
Then there is an extra ``impurity" level  
${\epsilon_1 \approx \gamma_1 \approx \exp[(s+\sigma)/2]}$ 
that is located above the continuum of extended modes. 
The contribution of this impurity to $V(\epsilon_s)$ 
is $\ln(\epsilon_s-\gamma_1)$. 
The contribution of the continuum can be neglected 
due to the Thouless relation. 
Hence the condition ${V(\epsilon_s)>V(0)}$ 
implies $s_c \approx \sigma/N$. 
\\     
}

\sect{Finding $S_c$ due to a weak-link}
We consider a clean ring of length ${L=Na}$ with lattice spacing $a$ 
and identical bonds (${w_n=1}$).  
We change one bond into a weak link (${w_1 \ll 1}$). 
This setup can be treated exactly in the continuum limit, 
where \Eq{e1} corresponds to a diffusion equation  
with coefficient ${D_0=wa^2}$ and drift velocity ${v_0=sD_0}$.  
The weak link corresponds to a segment 
where the diffusion coefficient is ${D_1 \ll D_0}$.  
Using transfer matrix methods we find the characteristic equation 
\be{50}
\cos(k) \ + \ \frac{1}{g} \left[\frac{k^2+(S_{\circlearrowleft}/2)^2}{2k}\right]\sin(k)
\ \ = \ \ \cosh \left(\frac{S_{\circlearrowleft}}{2} \right)
\eeq
where ${k^2= (L^2/D_0) z - (S_{\circlearrowleft}/2)^2}$, 
and $g=(D_1/D_0)/(a/L)$. We have taken here 
the limit ${a\rightarrow0}$, keeping~$(L,g,S_{\circlearrowleft})$ constant. 
The equation is graphically illustrated in \Fig{figReconstruction}. 
All the roots are real solutions provided the envelope 
of the left-hand-side (LHS) lays above the right-hand-side (RHS).
The minimum of the envelope of the LHS is obtained at ${z = S_{\circlearrowleft}^2/2}$.
Consequently we find that the threshold~$S_c$ obeys 
\be{23}
\frac{S_{\circlearrowleft}}{2g} \ \ = \ \ \cosh\left[ \frac{S_{\circlearrowleft}}{2} \right]
\eeq
provided $S_{\circlearrowleft} \gg g$, which is self-justified for small~$g$.   
The solution is given in terms of the Lambert function, 
namely ${S_c = -2 \mathbb{W}(-g/2)}$, which determines ${s_c=S_c/N}$.

The characteristic, equation \Eq{e50}, parallels the discrete version \Eq{e20}, 
with a small twist that we would like to point out.
Naively one would like to identify $\ln [2 (\text{LHS} -1)]$, 
up to a constant, with $\sum_{k=1}^{\infty} \ln(\epsilon-\epsilon_k)$, 
where the $\epsilon_k$ are the roots of \Eq{e50} with $S_{\circlearrowleft}{=}0$ 
in the RHS. This is tested in \Fig{figReconstruction}, 
and we see that there is a problem. Then one realizes 
that in fact an additional $k=0$ term with ${0 \lesssim \epsilon_0 < \epsilon_s}$ 
is missing. Going back to the discrete version it corresponds 
to an impurity-level that is associated with a mode which is located 
at the weak-link.
While taking the limit ${a\rightarrow0}$ this level becomes excluded.
Adding it back we we see that the agreement between \Eq{e20}      
and \Eq{e50} is restored. The residual systematic error as $k$ becomes 
larger is due to finite truncation of the number of roots used in the reconstruction. 
Making the approximation ${\ln(\epsilon_s-\epsilon_0) \approx \ln[(s/2)^2]}$,  
and noting that ${g\propto N}$, it is verified that the 
equation  ${V(\epsilon_s) = V(0)}$  for the complexity threshold 
is consistent with \Eq{e23}. \\

\sect{Finding $S_c$ for resistor-network disorder}
In the absence of stochastic field disorder, considering a percolating ring with ${\alpha>1}$, 
the threshold for complexity cannot be determined by the condition ${V'(0)<0}$ with \Eq{e19}, 
because for ${\mu_{\alpha}=1/2}$ we get formally ${V'(0)=0}$.  
Rather the threshold for complexity is determined by the condition ${V(0+)<V(0)}$, 
where $V(0+)$ is the the value of of~$V(\epsilon)$ 
in the vicinity of ${\epsilon\sim\epsilon_1}$. 
Recall that by the Thouless expression $(1/N)V(\epsilon)$
has been identified as the inverse localization length \cite{Shnerb1}.
We are dealing here with a ``conservative matrix" where the localization 
diverges at the potential floor as in the Debye model. 
It is well known that in the Debye model $(1/N)V(\epsilon) \propto \omega^2$ 
where $\omega=\sqrt{\lambda}$ corresponds to the frequency of the phonons.
Setting ${\omega_1 \propto 1/N}$ and ${V(0)\approx (S/2)^2}$ 
we conclude that ${S_c \propto 1/N^{1/2}}$. \\




\hidea{

\ \\ 

\section*{Acknowledgements}

This research has been supported by the Israel Science Foundation (grant No. 29/11).

\section*{Author contributions statement}

Both authors have contributed to this article. The numerical analysis and the figures have been prepared by DH, 
while the text of Ms has been discussed, written and iterated jointly by DC and DH.

\section*{Additional information}

The authors declare that they have no competing financial interests.
Correspondence and requests for materials should be addressed to DC [dcohen@bgu.ac.il] or to DH [hurowits@bgu.ac.il].

}

\clearpage
\end{document}